\documentclass[doublelinespacing]{elsart}
\usepackage{graphicx}

\usepackage{epsfig}

\usepackage{amssymb}

\begin{document}

\begin{frontmatter}

\title{Transport and cooling of singly-charged 
noble gas ion beams}

\author[lpc]{G. Ban,}
\author[lpc]{G. Darius,}
\author[lpc]{D. Durand,}
\author[lpc]{X. Fl\'echard,}
\author[lpc]{M. Herbane,}
\author[lpc]{M. Labalme,}
\author[lpc]{E. Li\'enard,}
\author[lpc]{F. Mauger,}
\author[lpc]{O. Naviliat-Cuncic,} 
\author[csnsm]{C. Guenault,}
\author[csnsm]{C. Bachelet,}
\author[cern]{P. Delahaye,}
\author[cern]{A. Kellerbauer,}
\author[ganil]{L. Maunoury,}
\author[ganil]{J.Y. Pacquet}
	
\address[lpc]{Laboratoire de Physique Corpusculaire, Caen, France}
\address[csnsm]{CSNSM, Orsay, France}
\address[cern]{CERN-ISOLDE, Geneva, Switzerland}
\address[ganil]{LIMBE-GANIL, Caen, France}

\begin{abstract}
The transport and cooling of noble gas singly-charged ion beams by means of 
a Radio Frequency Quadrupole 
Cooler Buncher (RFQCB) have been studied at the LIMBE low energy beam line  
of the GANIL facility. Ions as light as $^{4}He^+$ have been cooled and stored before
their extraction in bunches using $H_2$ as buffer gas. 
Bunches characteristics have been studied as a function of the
parameters of the device. Sizeable transmissions of up to 10 $\% $ have been 
obtained. A detailed study of 
the lifetime of ions inside the buncher has been performed giving an 
estimate of the charge exchange
cross-section.
Results of a microscopic Monte-Carlo transport code show reasonable agreement with 
experimental data.     
\end{abstract}

\end{frontmatter}

\section{Introduction}
Beam handling is nowadays a very active field of research in nuclear physics. In this context, 
Radio Frequency Quadrupole Cooler-Buncher (RFQCB) devices have shown to play an increasing role 
to manipulate radioactive beams near low energy 
facilities \cite{vandenberg,herfurth,nieminen,maier,savard,liu,rodriguez}. 
The Laboratoire de Physique Corpusculaire (LPC) in Caen  
has developped a RFQCB to cool and guide exotic ions 
produced near the SPIRAL facility at GANIL \cite{spiral}. 
One of the planned experiments will focus on the precise measurement of the 
$\beta-\nu$ angular correlation parameter in the $\beta$ decay of $^6He$
\cite{lipari}. 
Such a measurement allows to 
search for the possible 
existence of exotic couplings not allowed by the Standard Model \cite{hergzeg}. 
The experiment will be performed inside a transparent Paul trap \cite{delahaye}. 
Unstable $^6He^+$ ions will be confined under high vacuum in a few $mm^3$ 
volume in the center of the trap. Following their $\beta$ decay, coincidences 
between the $\beta$ and the recoil ion, $^6Li$, will be recorded.
The radioactive ions will be provided by the LIRAT \cite{auger2,auger3} beam line, a low 
energy radioactive isotope beamline 
at GANIL. It uses an Electron Cyclotron Resonance Ion Source 
for ion production after fragmentation reactions \cite{maunoury1}.
Because of the large temperature and electron density 
inside an ECR plasma, the ions are heated, 
and their temperature is in the few $eV$ range inducing large 
emittances, up to 100 $\pi mm.mrad$ for ion kinetic energies near 30 $keV$.
The use of a RFQCB filled with a buffer gas is thus intended to prepare the beam before 
injection inside the Paul trap in order to allow the trapping by 
reducing the initial emittance.
The principle is to confine radially the ions by means of a radio frequency 
quadrupole field and to slow down their motion by collisions 
with the atoms or molecules of the so-called buffer gas. 
An additional electrostatic longitudinal 
field drives the ions towards the end of the device where 
they can be stored in a potential well and extracted as a bunch. They can then  
be efficiently injected in the Paul trap. 
We have performed an experiment using stable
singly-charged ions produced by the LIMBE low energy beam line at GANIL \cite{maunoury2}.  
After a description of
the apparatus, the results concerning the cooling and 
bunching of $^4He^+$ and $^{40}Ar^+$ ions using $He$ and $H_2$ as buffer gases 
are presented. They are
compared with the outcome of a microscopic transport model.  
\section{The experiment}
\subsection{Experimental set-up}
\begin{figure}[!htb]  
\begin{center}
\psfig{file=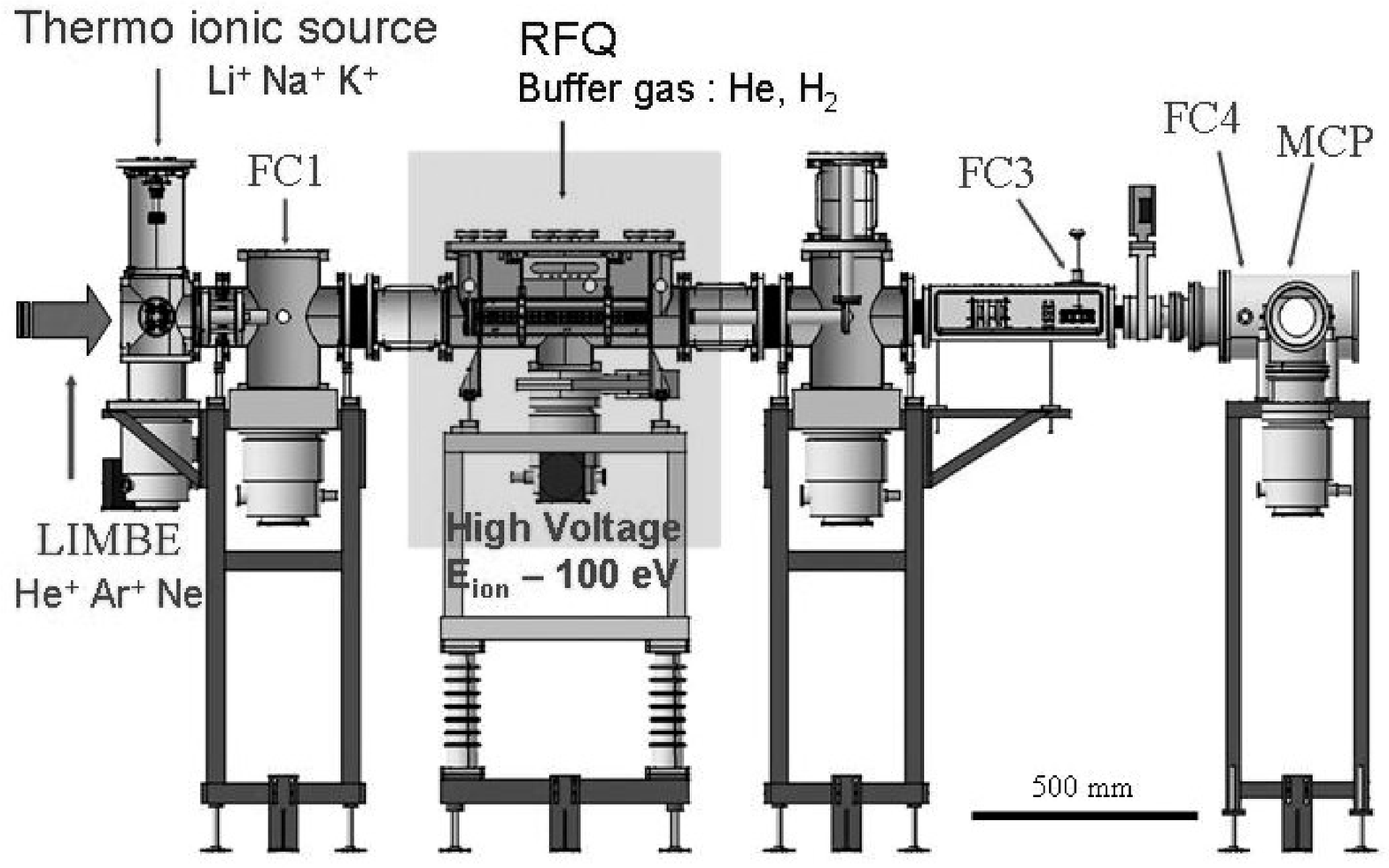,width=140mm,height=100mm}
\caption{\it Schematic view of the experimental set-up. }  
\label{setup}
\end{center}
\end{figure}

The experimental set-up is displayed in Fig.\ref{setup}. 
Ion beams come from the left and enter
the system passing through 
a first Faraday Cup, FC1, used for monitoring purposes. 
Note also the
presence of a thermo-ionic source perpendicular to the beam axis which can be
used to produce alkaline ions. Data obtained for such ion beams will be
discussed elsewhere \cite{alkaline}. Ions are then transported and focussed 
at the entrance
of the RFQCB. This latter is placed on a high voltage 
platform whose voltage is 100 $V$ lower than the extraction voltage of the ECR source. Therefore,
incoming ions have kinetic energies close to 100 $eV$ with respect to the RFQCB.  
Typical energies of singly-charged ions 
delivered 
by the LIMBE facility are a few $keV$ \cite{maunoury2}. The RFQCB is described in details
elsewhere \cite{rfqcb}. Since the mass of the constituents of the buffer gas must be
lighter than the one of the ions, two light gases have been used: $H_2$ and the most
commonly used $He$. The reason why $H_2$ had to be used 
in these tests for $^4He^+$ ions is due to the strong charge exchange
resonant process between $He$ atoms and singly-charged $He^+$ ions. This is also the
reason why the
future experiment involving $^6He$ ions will use $H_2$. 
Ions are stored at the end of the RFQCB inside the buncher. After a variable
time (hereafter called the "cooling" time, $t_{cool}$), ions are ejected by
switching  the configuration of the electrostatic potential as shown in Fig.
\ref{field}.
After extraction and passage thbrough a pulse-down electrode, the ion 
kinetic energy is close to 1~$keV$. The rest of 
the line is dedicated to the
transport and the diagnostics (FC2) of the cooled beam up to 
the detection device 
consisting of a micro-channel plate (MCP). 
\subsection{Operating modes}
\label{mode}
Measurements have been carried out to observe the cooling
and bunching of ions inside the RFQCB while varying the parameters of the device.
Measurements were firstly performed with a continuous beam. Faraday cups were then used to 
measure beam currents at the entrance (FC1) and at the exit (FC2) of the RFQCB thus allowing to estimate 
the transmission through the device.    
However, in order to have access to the time dependence of the whole process,
the ion beam delivered 
by LIMBE could be pulsed by applying a bias between two parallel
plates located along the beam line with an adjustable duration between 1 and 5 $ms$, 
and a frequency between 10 and 100 $Hz$. Therefore, ions were swept away from 
the beam line 
when the bias was on and were transported towards the entrance of the RFQCB when the bias was 
off. Let $t_0$, be
the instant when the beam passes through the plates. Since the time of flight between the
deflecting plates and the entrance of the RFQCB is a constant for a given ion, $t_0$ can be safely 
used as a reference "start"
for the study of the time dependance of the cooling and bunching process.
The "cooling" time, $t_{cool}$, could be varied between 0 and 50 $ms$ 
after $t_0$. Short values of $t_{cool}$ (less than a $ms$) 
allow the study of the
diffusion and transport of the incoming ions through the cooler-buncher as discussed in
section \ref{seccool} while larger
values give access to the lifetime of the ions inside the buncher (see section \ref{life}).  
\subsection{Brief description of the transport model used for the simulations}
A simulation tool has been developped to design the characteristics of the device and to compare with
experimental data.
It is described in details elsewhere \cite{lipari,montecarlo}. 
This is a Monte 
Carlo code describing at the microscopic level the collisions between the incoming 
ions and the buffer gas. The kinematics and the rate of the collisions
are estimated using interaction potentials as realistic as possible. These latter are  
"tested" by comparison of the results of the code with experimental mobilities 
and diffusivities \cite{ellis1,ellis2,ellis3,viehland}. The transport of the ions in the 
electromagnetic environment 
is considered so that the total transmission process of the ions inside the 
cooler-buncher can be simulated including the storage inside the buncher.
\section{Measurements}
Since it was not possible to obtain either absolute calibrations of the FC's or the absolute
efficiency of the MCP with sufficient accuracy, we have decided to normalize the experimental
data to the results of the simulation. 
However, the transmissions obtained by measuring currents on the Faraday cups or by
counting the ions on the MCP are comparable (within a factor 2) 
with the transmissions predicted by the Monte Carlo code.  
\subsection{Transmission: influence of the buffer gas pressure}
\begin{figure}[!htb]
\begin{center}
\epsfig{file=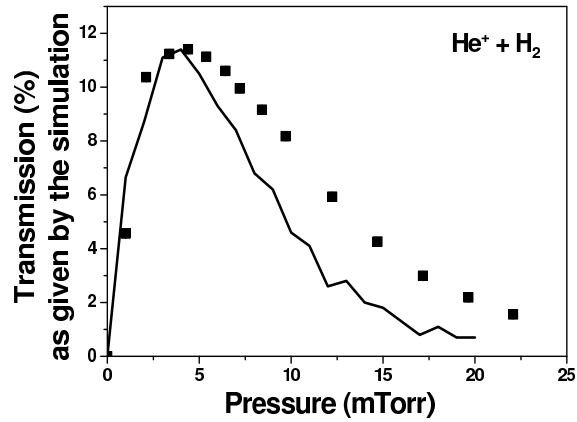,width=80mm,height=60mm}
\caption{\it Transmission as a function of the pressure for $^4He^+$ ions 
extracted from the RFQCB using $H_2$ as buffer gas. Experimental data (black
squares) have been
normalized to the calculation (solid line) at the maximum value.}  
\label{pres1}
\end{center}
\end{figure}

\begin{figure}[!htb]
\begin{center}
\epsfig{file=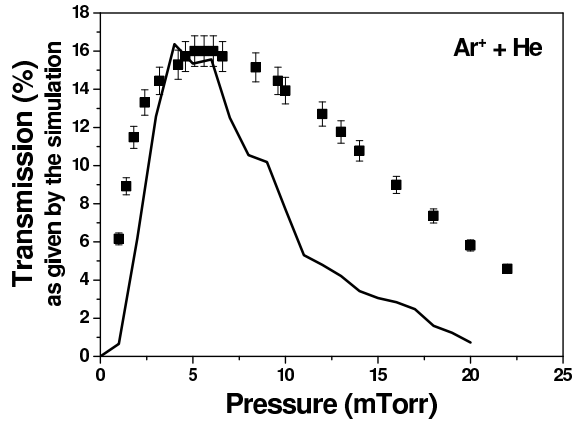,width=80mm,height=60mm}
\caption{\it Same as Fig.\ref{pres1} but for $^{40}Ar^+$ ions and $He$ 
as buffer gas.}  
\label{pres2}
\end{center}
\end{figure}

We first address the evolution of the transmission through the RFQCB as a function 
of the buffer gas pressure. Fig. \ref{pres1} and \ref{pres2} show experimental results (black squares) measured
in the continuous mode and the results of the calculation (lines) for resp. $^4He^+$ in $H_2$ and $^{40}Ar^+$ in
$He$. Both distributions exhibit a maximum around 4-5 $mTorr$ for
$^4He^++H_2$ and 5-6 $mTorr$ for $^{40}Ar^++He$. The transmission is low at low pressure because 
there are not enough collisions between ions and
the buffer gas to properly guide the ions towards the exit of the buncher. 
As the pressure increases, the collision rate increases and hence the transmission.
However, for large values of the pressure, the
transmission decreases because there are then too many collisions so that ions
are lost in the structure near the exit of the device. In other words, the size of the ion cloud is larger than
the exit diameter of the apparatus. 
As far as calculations
are concerned, although the maximum value is correctly reproduced, the
distribution is much too narrow as compared to experimental data. A possible
explanation lies in the fact that a constant homogeneous pressure has been  
assumed inside the RFQCB for the simulations. In principle, one should take into account 
a possible gradient of the pressure near 
the exit (and also the entrance) of the apparatus due to gas flow. This has not yet been implemented in the
simulation.  
\subsection{Transmission: influence of the RF parameters}
\begin{figure}[!htb]
\begin{center}
\epsfig{file=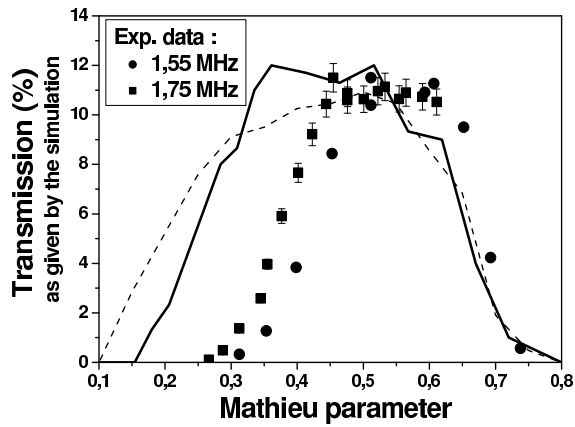,width=80mm,height=60mm}
\caption{\it Transmission as a function of the Mathieu parameter 
for $^4He^+$ and $H_2$ as buffer gas. The two data sets correspond to 
two different values of the frequency $f$. For each fixed value of the frequency, the bias $V_0$
applied on the rods of the RFQCB was changed in order to scan the range of interest of the $q$
parameter. Calculations: solid line ($f$=1.55 MHz), dashed line ($f$=1.75 MHz).}  
\label{mathieu}
\end{center}
\end{figure}

An exploration of the stability region has been undertaken 
by varying the Mathieu parameter, $q$, defined by:
\begin{equation}
q=\frac{4eV_0}{mr^2_0\omega^2}
\end{equation}
where $e$ is the charge of the ion, $V_0$ the potential applied to
the rods of the quadrupole, $m$ the ion mass, $\omega$ the radio-frequency and
$r_0$ the radius of the quadrupole. In the case of $^4He^+$+$H_2$, typical values are $V_0$=100 $V$ and
$\nu=\omega/2\pi$ between 1 and 2 $MHz$. As shown in Fig. \ref{mathieu}, the maximum transmission is obtained
(whatever the considered frequency) around $q$=0.5. This is in rough agreement with the calculation although this
latter predicts a much larger range of optimal transmission starting around $q$=0.3-0.4: a fact that is not
observed in the experimental data. A possible explanation lies on the distorsion of the RF outputs (which were
not fully sinusoidal) for low values of the applied bias, $V_0$, corresponding to low values of $q$. 
\subsection{Optimized transmissions}
In the continuous beam mode, transmissions have been estimated by comparing currents measured on
the different FC's.  
There are uncertainties concerning these figures because the first FC (FC1) could not be placed 
close enough 
to the entrance of the RFQCB so that the flux measured by FC1 may be overestimated 
with respect to the
actual
flux entering the apparatus. More, as already mentionned, the FC's have not been calibrated 
on absolute values. We therefore 
estimate the uncertainty in the transmission measurement around a
factor 2. A compilation is shown in Table \ref{table1}. 
For all systems, a systematic exploration of the influence of the pressure and of the RF parameters 
has been performed and the numbers quoted in Table \ref{table1} correspond to 
optimal values. One should
however note that for technical reasons, the $q$ parameter could not be set to its 
optimum value (around 0.5)
for all studied systems.  
Note also that the transmission for
$^{40}Ar^+$ + $H_2$ is zero because of a strong charge exchange resonant effect due to the similarity of the 
values of the ionization
potentials of these two species: 15.755 and 15.427 $eV$ respectively. Apart from that, transmissions 
between 5 and 10
$\%$ have been obtained. No significant differences were observed for $^{22}Ne^+$+$He$ 
and $^{22}Ne^+$+$H_2$. As stated above,     
the measured transmissions are within the expectations of the simulation.
\subsection{Time-of-flight measurement}
\label{tof}
\begin{figure}[!htb]
\begin{center}
\epsfig{file=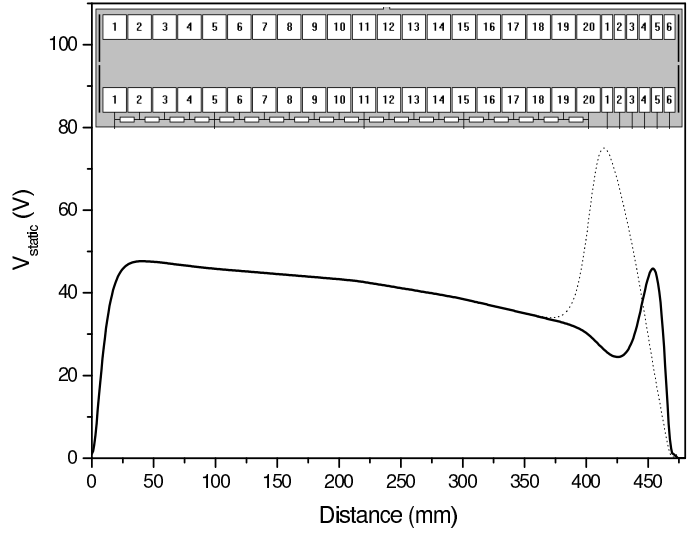,width=90mm,height=70mm}
\caption{\it Electrostatic field along the symmetry axis 
as a function of the distance in
the RFQCB for two different configurations. Configuration {\it "storage"} (thick line) 
allows to confine ions in a potential well inside the buncher while the configuration 
{\it "extraction"} (dotted line)
pushes the ions outside the buncher. The last configuration is obtained by switching the voltage 
applied to 
the buncher electrodes. The geometry of the segmented electrodes constituting the 
cooler (labelled from 1 to
20 starting from the left) and the buncher (labelled from 1 to 6) is shown in the upper part of
 the figure. The values of the expected electric field with such a geometry 
have been obtained 
using the SIMION \cite{dahl}
software computer code. Potentials have been calculated with respect
to the potential of the
platform on which the device is placed.}  
\label{field}
\end{center}
\end{figure}

\begin{figure}[!htb]
\begin{center}
\epsfig{file=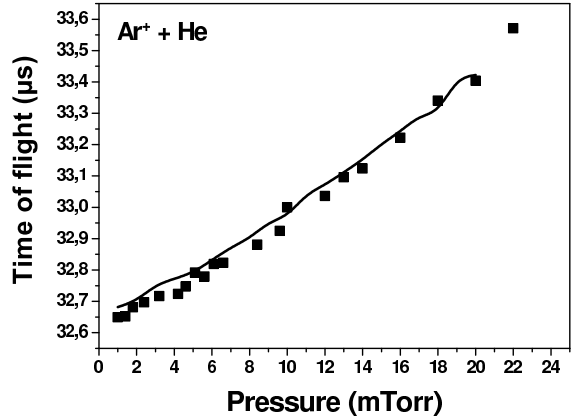,width=80mm,height=60mm}
\caption{\it Mean time-of-flight distribution for $^{40}Ar^+$ ions after
extraction from the RFQCB using $He$ as buffer gas as a function of pressure.
Black squares: experimental data, solid line: results of the simulation.}  
\label{arhe2}
\end{center}
\end{figure}

\begin{figure}[!htb]
\begin{center}
\epsfig{file=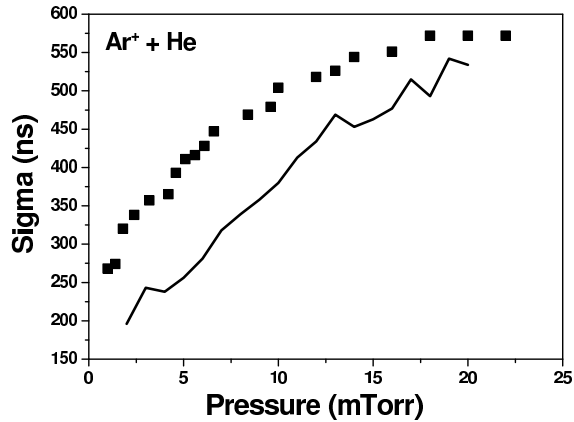,width=80mm,height=60mm}
\caption{\it Width of the time-of-flight
 distribution under the same conditions than Fig.\ref{arhe2}.}  
\label{arhe3}
\end{center}
\end{figure}

\begin{figure}[!htb]
\begin{center}
\epsfig{file=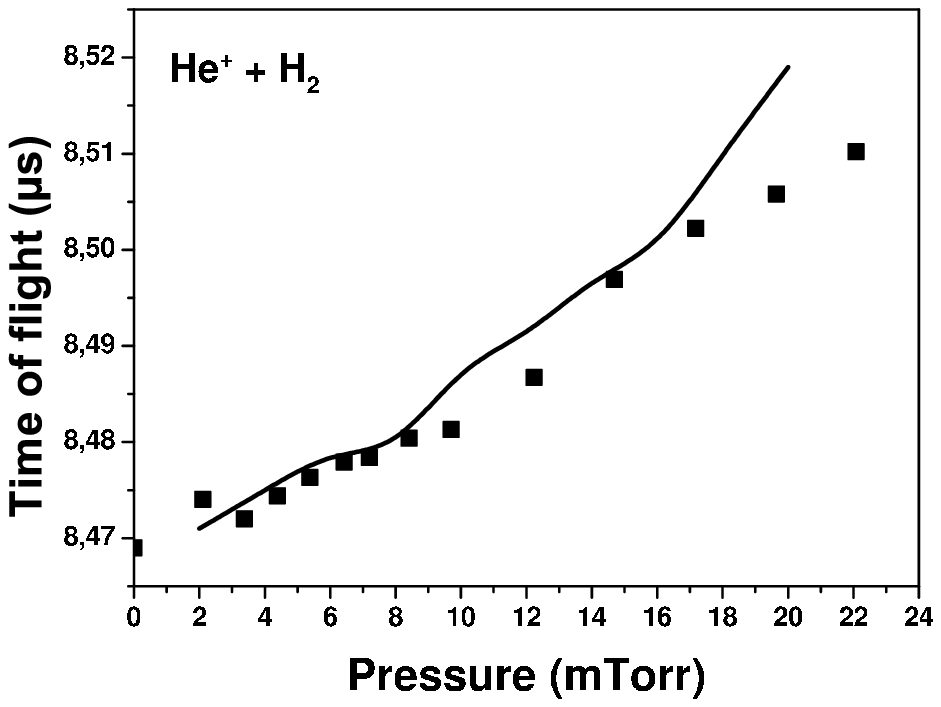,width=80mm,height=60mm}
\caption{\it Same as Fig.\ref{arhe2} but for $^4He^+$ ions and $H_2$ as buffer gas.}  
\label{heh21}
\end{center}
\end{figure}

\begin{figure}[!htb]
\begin{center}
\epsfig{file=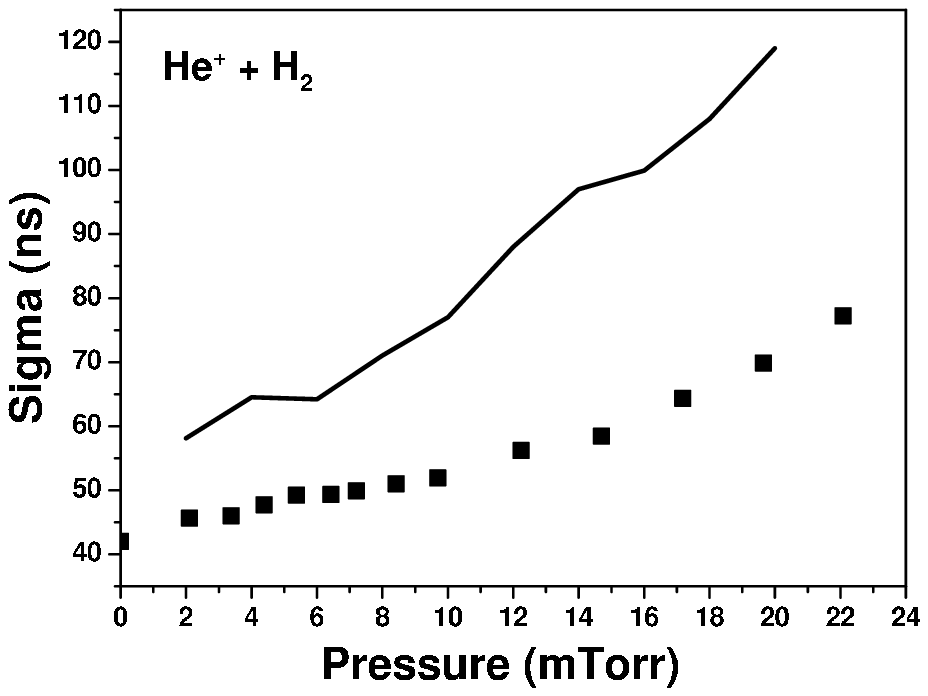,width=80mm,height=60mm}
\caption{\it Same as Fig.\ref{arhe3} but for $^4He^+$ ions and $H_2$ as buffer gas.}  
\label{heh22}
\end{center}
\end{figure}

We now consider the time-of-flight (ToF) 
distributions of the extracted 
bunches. This is achieved by measuring
the time of arrival of the ions on the MCP detector with respect 
to the time associated with the switch in the potential configuration of the buncher as shown in Fig.
\ref{field}. 
We first discuss the $^{40}Ar^+$+$He$ case in Fig. \ref{arhe2} and \ref{arhe3}.
In Fig. \ref{arhe2}, the evolution of the mean value of the time-of-flight is shown as a function of the
pressure for experimental data (black squares) and for simulated data (solid line).
As expected, an increase of the ToF is observed as pressure increases. This is understood as the growth
of collision rate near the exit of the buncher. In conjunction, the 
pressure has also an
influence on the ToF width.  
Most of the time distribution width is due to the finite extent of the confined cloud of ions
inside the buncher. Indeed, the electrostatic potential has such a structure 
(see Fig. \ref{field}) that ions are located in
a region of the order of a few $mm$. When the electrostatic barrier is lowered
down, those ions which are located close to the exit have a shorter travel inside the buncher and 
suffer less residual
collisions with the gas than those located more inside the buncher. Hence, they 
have a shorter diffusion time towards the exit and this leads to a spread both in time and space 
of the bunches. These effects are correctly accounted for by the simulation although the model somehow
underestimates the time width. 
The agreement is not so good when considering the $^4He^+$+$H_2$ case (Fig. \ref{heh21} and 
\ref{heh22}). In particular, the width is there strongly overestimated by the simulation. One should note
that the interaction potential for such a system is poorly known. In particular, for this system, 
it was not possible to
confront the results of the model with experimental mobilities and/or diffusivities since these latter
are not available. Therefore, the parameters of the potential were taken to be those of $^6Li^+$.    
\subsection{Measurement of the "cooling" time: $t_{cool}$}
\label{seccool}
\begin{figure}[!htb]
\begin{center}
\epsfig{file=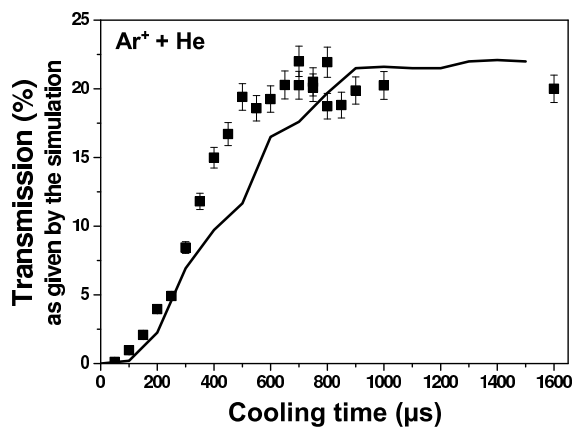,width=80mm,height=60mm}
\caption{\it Transmission as a function of the 'cooling' time defined in the
text. The ion is $^{40}Ar^+$ and the buffer gas is $He$. Black squares: experimental data, solid line: calculation.}  
\label{tcool1}
\end{center}
\end{figure}

\begin{figure}[!htb]
\begin{center}
\epsfig{file=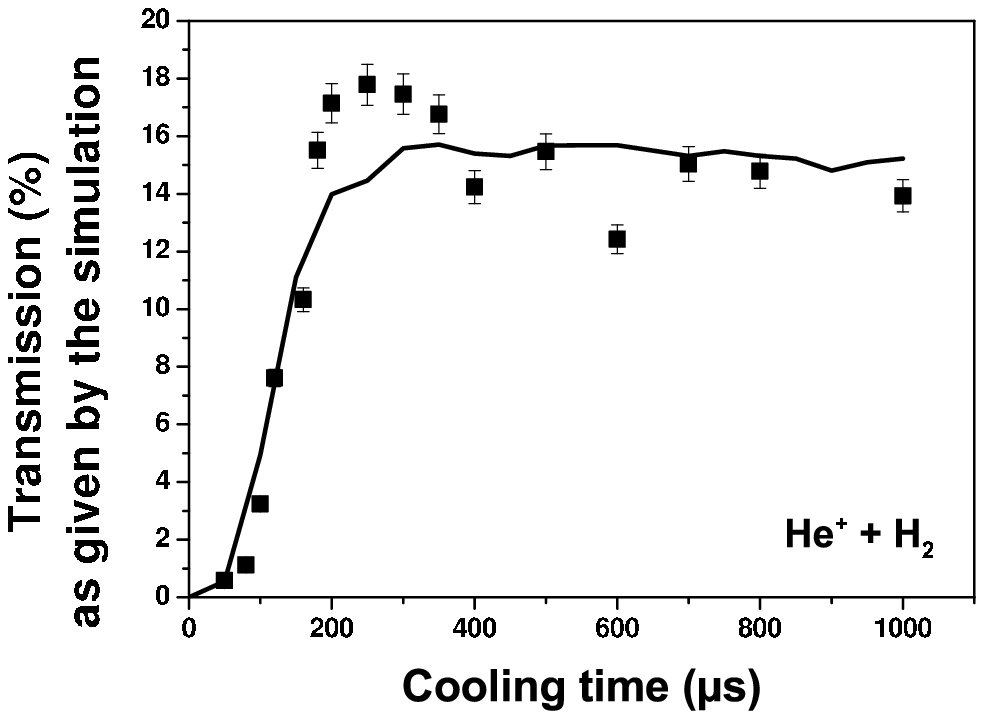,width=80mm,height=60mm}
\caption{\it Same as Fig.\ref{tcool1} but for $^4He^+$ ions and $H_2$ as buffer
gas.}  
\label{tcool2}
\end{center}
\end{figure}

In order to understand in more details the transport properties ot the device, we have
performed measurements of the time dependance of the transmission factor. To this end,  
the beam was pulsed as described previously in section \ref{mode}.
A short pulse was injected in the RFQCB and the extracted ions
were detected and counted as a function of $t_{cool}$. 
For short values, this is a measure of 
the time needed for the ions to travel   
through the device before reaching the buncher. This quantity is mainly
governed by the diffusion process of the ions throughout the gas and is therefore an
interesting test as far as the validity of the simulations is concerned. 
Figs \ref{tcool1} and \ref{tcool2} show the results of the comparisons. For very short 
values of $t_{cool}$, ions
have no time to reach the end of the device and there is no signal. As soon as
typical values are applied of the order of a 
few hundreds $\mu s$, the transmission rapidly increases and then  
saturates. Here, those ions which have survived their
travel across the apparatus have all reached the buncher. 
A comparison of the results of the simulation with the
data shows a rather good agreement although for the $^{40}Ar^+$+$He$ system, the
saturation value of $t_{cool}$ is slightly overestimated.   
It is worth noting that in order to obtain an experimental 
"clean" signal, it was found necessary to 
empty the cooler before each injection by extracting those ions from the previous 
injection which remained in the device.
\subsection{Lifetimes}
\label{life}
\begin{figure}[!htb]
\begin{center}
\epsfig{file=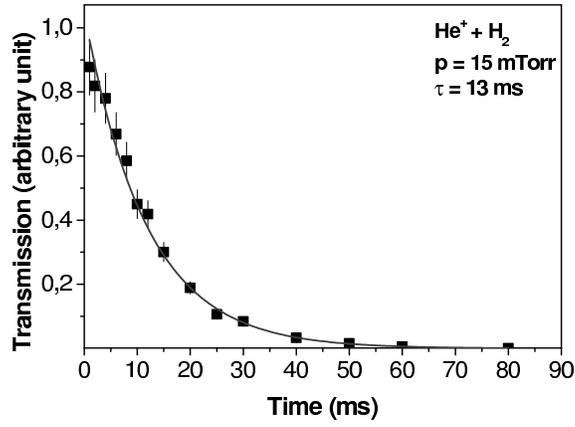,width=80mm,height=60mm}
\caption{\it Lifetime measurement: transmission as a function of $t_{cool}$
for $^4He^+$. The line corresponds to an exponential decay fit with a time
constant of 13 $ms$. Here, the $H_2$ pressure is 15 $mTorr$.}  
\label{lifetime1}
\end{center}
\end{figure}

\begin{figure}[!htb]
\begin{center}
\epsfig{file=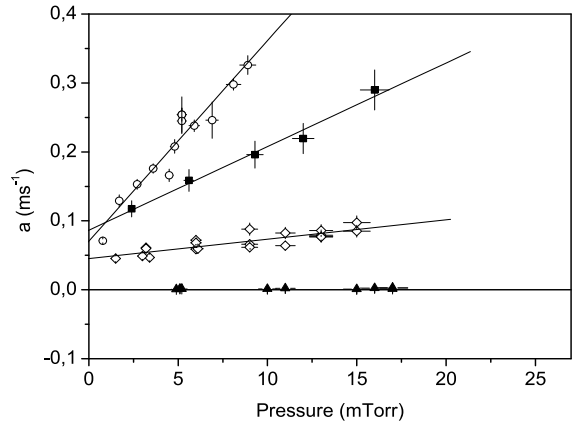,width=80mm,height=60mm}
\caption{\it Evolution of $a=1/\tau$ (in $ms^{-1}$)
as a function of the pressure for various
combinations of ion-buffer gas. Solid lines are the result of a linear 
fit according to Eq. \ref{eq2}. Black triangles: $Li^+$+$H_2$, black squares: $^{22}Ne^+$+$H_2$, open
circles: $Li^+$+$He$, open squares: $^4He^+$+$H_2$.}   
\label{lifetime2}
\end{center}
\end{figure}

Extending the time measurement described in the previous section well above the $ms$ regime allows to address the question of the lifetime of
the ions stored inside the buncher. Measurements
of the transmission have been performed up to 90 $ms$ as shown in Fig. \ref{lifetime1}.
For each pressure, an exponential fall-off in the time distribution has been observed and measured. 
A compilation of the results obtained for different combinations of ion-buffer gas is displayed in
Fig. \ref{lifetime2}. The discussion of alkaline ions will be 
performed elsewhere: they are here only for a matter of comparison. 
The lifetime $\tau$ reads:
\begin{equation}
\label{eq1}
\tau=\frac{1}{\rho <\sigma_{CT} v_{rel}>}
\end{equation}
where $\rho$ is the buffer gas density and $<\sigma_{CT} v_{rel}>$ is the average product of the 
charge transfer cross-section and of the relative velocity between the ions and the buffer 
gas atoms or molecules. One then obtains: 
\begin{equation}
\label{eq2}
\frac{1}{\tau}\propto p <\sigma_{CT} v_{rel}>
\end{equation}
The inverse of the lifetime is proportionnal to the buffer gas pressure if the loss of ions 
is only due to charge
exchange reactions. Using Eq. \ref{eq2}, we have access to the
product $<\sigma_{CT} v_{rel}>$. In the case of
$^4He^++H_2$, one obtains experimentally: $\tau p \simeq$ 200 $ms.mTorr$ which leads to $<\sigma_{CT} v_{rel}>
\simeq$ 1.4 $10^{-13} cm^3/s$. This value is in good agreement with the value quoted in \cite{kimura}.
However, one should notice in Fig. \ref{lifetime2} that the experimental values of the
inverse of the lifetime do not cross the origin when the pressure is zero. 
We believe that the deviations observed are essentially due to uncertainties in the absolute values of the
pressure and also probably to the presence of polluants such as water or $N_2$ inside the buncher. 
\section{Conclusions}
In this paper, we have shown the results of an experiment aiming at cooling and bunching
singly-charged light noble gas ions by means of a RFQCB installed on the beam line LIMBE
at the GANIL facility. 
The
following results have been obtained:
\begin{itemize}
\item
Ions as light as $^4He^+$ have been cooled and bunched using $H_2$ as buffer gas. The total 
transmission through 
the device has been found of the order of 10 $\%$.  
\item 
The "cooling" time, that is the time for the incoming ions to reach the bunching 
section of the device, has been
 found to be of the 
order of 300 $\mu s$ for $^4He^++H_2$ and 800 $\mu s$ for $^{40}Ar^++He$. Bunches of cooled $^4He^+$ ions have been
produced with a time structure width of the order of 50 $ns$.  
\item 
Lifetime of the ions inside the buncher were measured by observing the fall-off of 
the transmission as a function of time up 
to a few tens of
$ms$. A loss of ions in the case of $^4He^+$ has been observed due to charge transfer 
reactions. The value of the cross-section for charge transfer deduced from the evolution of the
lifetime as a function of the buffer gas pressure is in agreement with previous measurements.  
\item
Experimental data   
are in reasonnable agreement with the results of a microscopic Monte Carlo transport
code. 
\end{itemize}

The authors would like to thank warmly the technical 
staff members of LPC CAEN for help in the preparation of and during the experiment, in
particular Ph. Desrues, Y. Merrer and C. Van Damme for the mechanical part of the
experiment and J. Br\'egeault and Ph. Vallerand for the electronics. 

Part of the set-up has been funded by the Region Basse-Normandie.
This work was achieved and is pursued within the NIPNET Collaboration under contract HPRI-CT-2001-50034. 
   

\newpage

\begin{table}
\begin{center}
\begin{tabular}{|l|l|l|}
\hline
\tt Ion & Buffer Gas & T(\%) \\
\hline
\tt $^4He^+$ & $H_2$ & 5-10 \\
\hline
\tt $^{22}Ne^+$ & $He$,$H_2$ & 5-10 \\
\hline
\tt $^{40}Ar^+$ & $He$ & 5-10 \\
\hline
\tt $^{40}Ar^+$ & $H_2$ & 0 \\
\hline
\end{tabular}
\caption {\it Compilation of the transmissions obtained for the different pairs of
ion-buffer gas constituents studied in this work.}
\label{table1}
\end{center}
\end{table}
\end{document}